\documentclass[aps,prl,showpacs,twocolumn,floatfix]{revtex4-1}
\usepackage{eurosym}
\usepackage{amsfonts}
\usepackage{mathrsfs}
\usepackage{color}
\usepackage{xspace}
\usepackage{subfigure}
\usepackage{longtable}
\usepackage{xspace}
\usepackage{multirow}
\usepackage{tabularx}
\usepackage{amsmath}
\usepackage{amssymb}
\usepackage{graphicx}
\usepackage{dcolumn}
\usepackage{bm}
\usepackage[colorlinks,urlcolor=blue,citecolor=blue,linkcolor=blue]{hyperref}

\begin{document}

\title{Fulde-Ferrell Superfluids without Spin Imbalance in Driven Optical Lattices}
\author{Zhen Zheng$^{1,2}$}
\author{Chunlei Qu$^{1}$}
\author{Xubo Zou$^{2}$}
\author{Chuanwei Zhang$^{1}$}
\thanks{chuanwei.zhang@utdallas.edu}

\begin{abstract}
Spin-imbalanced ultracold Fermi gases have been widely studied recently as
a platform for exploring the long-sought Fulde-Ferrell-Larkin-Ovchinnikov
superfluid phases, but so far conclusive evidence has not been found.
Here we propose to realize an Fulde-Ferrell (FF) superfluid without spin imbalance in a
three-dimensional fermionic cold atom optical lattice, where $s$- and $%
p $-orbital bands of the lattice are coupled by another weak moving optical
lattice. Such coupling leads to a spin-independent asymmetric Fermi surface,
which, together with the $s$-wave scattering interaction between two spins,
yields an FF type of superfluid pairing. Unlike traditional schemes, our
proposal does not rely on the spin imbalance (or an equivalent Zeeman field)
to induce the Fermi surface mismatch and provides a completely new route for
realizing FF superfluids.
\end{abstract}
\affiliation{$^1$Department of Physics, The University of Texas at Dallas, Richardson,
Texas 75080, USA \\
$^2$Key Laboratory of Quantum Information,
and Synergetic Innovation Center of Quantum Information \& Quantum Physics,
University of Science and Technology of China, Hefei, Anhui 230026, People's Republic of China}
\maketitle

The Fulde-Ferrell-Larkin-Ovchinnikov (FFLO) state, characterized by Cooper
pairs with finite center-of-mass momenta~\cite{FF64,LO64}, is a central
concept for understanding many exotic phenomena in different physics
branches~\cite{Gasalbuoni2004}. A crucial ingredient for realizing FFLO
states is a large Zeeman field that induces a Fermi surface mismatch of two
paired spins \cite{FF64,LO64}. In recent years, FFLO states have been
extensively studied in ultracold Fermi gases, where the population
imbalance between two atomic internal states (pseudo-spins) serves as an
effective Zeeman field~\cite%
{Zwierlein2006,Hulet2006,Hu_pra2006,Parish2007,Liu2007,Torma2006,Machida2006,Koponen_njp2006,Hu_prl2007, time_flight,Liao2010}%
. Despite the intrinsic advantages of cold atoms compared to their solid
state counterparts, conclusive evidence of FFLO states has not been found
yet because of various obstacles. For instance, a large Zeeman field
suppresses the superfluid order parameter, leading to a very narrow
parameter region for FFLO states in 2D or 3D which can be easily destroyed
by thermodynamic fluctuations~\cite{Zwierlein2006,Hulet2006,Hu_pra2006}. In
1D, the parameter region for FFLO states could be large, but the quantum
fluctuation is strong \cite{Liao2010,Hu_prl2007,Liu2007}. The recently
proposed schemes using spin-orbit coupling and in-plane Zeeman field in a 3D
Fermi gas may potentially overcome these obstacles \cite%
{Zheng2013,Wu2013,Dong2013,topoFF1,topoFF2,topoFF3,topoFF4} in principle,
but they face practical experimental issues such as the large spontaneous
photon emission from the near-resonant Raman lasers \cite%
{Galitski2013,Williams2013,Fu2014,Ian2011,Pan,Peter,Peter2,Chen,Zhang2012,Martin2012}
and the strong three-body loss at Feshbach resonance in the presence of
spin-orbit coupling~\cite{Zhang2012,Martin2012,Williams2013,Fu2014}.

In this Letter, we propose a new route for realizing FF superfluids in
ultracold Fermi gases without involving population imbalance of two spin
states that interact for generating Cooper pairing. Instead, we induce an
asymmetric Fermi surface for the generation of FF states by other means and
the populations of the two spins are fully equal. Our main results are the
following:

1) We show that the \textit{s}- and \textit{p}$_{x}$-orbital bands of a 3D
static optical lattice can be coupled using a weak 1D moving optical lattice
along the $x$ direction, which can be generated by two counter-propagating
lasers with the frequency difference matching the \textit{s-p}$_{x}$ band
gap. The \textit{s}- and \textit{p}$_{x}$-bands can be denoted as the
band pseudospin, and the moving lattice induces a band-pseudospin-momentum
(i.e., spin-orbit) coupling and an in-plane Zeeman field, which yield an
asymmetric Fermi surface along the $x$ direction. The realization of such
band-pseudospin-momentum coupling may provide a new platform for exploring
exotic spin-orbit coupling physics.

2) We show that the asymmetric Fermi surface, together with the \textit{s}%
-wave pairing interaction between two equally populated hyperfine spin
states, can induce an FF type of Cooper pairing within a large parameter
region in the 3D optical lattice, in sharp contrast to the narrow parameter
region for the spin-imbalanced Fermi gas~\cite{Hu_pra2006,Liu2007}. Because
of the 3D nature of the FF superfluids, the quantum fluctuations are also
suppressed. The generated FF state is thermodynamically much more stable
than the spin-imbalanced Fermi gas. Compared to the spin-orbit coupled
schemes~\cite{Zheng2013,Wu2013,Dong2013,topoFF1,topoFF2,topoFF3,topoFF4}
that require near resonant Raman lasers \cite%
{Ian2011,Pan,Peter,Peter2,Chen,Zhang2012,Martin2012,Williams2013,Fu2014},
all lasers used here are far-detuned, therefore the proposed scheme should
work for all types of fermionic atoms, including $^{6}$Li~\cite{Liao2010}.
Furthermore, because the hyperfine spins are not coupled with the momentum,
the \textit{s}-wave scattering interaction should be the same as regular
Fermi gases without significant three-body loss at Feshbach resonance. These
intrinsic advantages of our spin-balanced scheme make it experimentally more
feasible than the spin-imbalanced schemes (with \cite%
{Zwierlein2006,Hulet2006,Hu_pra2006,Parish2007,Liu2007,Torma2006,Machida2006,Koponen_njp2006,Hu_prl2007, time_flight,Liao2010}
or without spin-orbit coupling \cite%
{Zheng2013,Wu2013,Dong2013,topoFF1,topoFF2,topoFF3,topoFF4}), and thus may
open a new route for observing FF superfluids.

\begin{figure}[t]
\centering\includegraphics[width=0.48\textwidth]{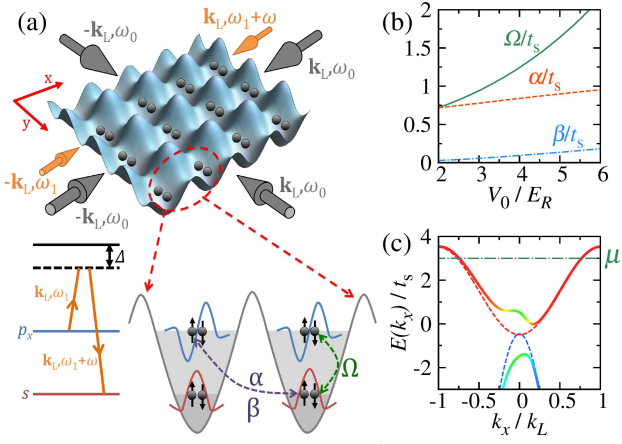}
\caption{(Color online) (a) An illustration of the experimental proposal: a
moving lattice (orange arrows) induces effective two-photon Raman couplings
between $s$- and $p_{x}$-bands of a 3D static optical lattice (only show 2D
here, gray arrows). (b) Plot of different coupling strengths $\Omega $, $%
\protect\alpha $ and $\protect\beta $ as a function of the static lattice
depth $V_{0}$. The moving lattice depth $V_{x}^{\prime }=0.8E_{R}$ with the
recoil energy $E_{R}=h^{2}/2ma^{2}$. (c) The single-particle band structure
for $V_{0}=3.0E_{R}$. $h=5.5t_{s}$ and $\protect\mu=3t_{s}$ with $%
t_{s}=0.111E_{R}$. The colors represent hybrid orbital compositions for each
momentum state (red for $s$- and blue for $p_{x}$-orbital states). Dashed
lines: bare $s$- and shifted $p_{x}$-orbital bands without coupling.}
\label{fig-illustrate}
\end{figure}

\textit{Asymmetric Fermi surface in a driven optical lattice}: Consider a
degenerate spin-$1/2$ Fermi gas trapped in a static 3D optical lattice. Our
proposed experimental setup is illustrated in Fig. \ref{fig-illustrate}(a).
An additional 1D moving lattice along the $x$ direction is applied to couple
the \textit{s}- and $p_{x}$-orbital bands of the static lattice. The moving
lattice is generated by two counter-propagating lasers with a frequency
difference of $\omega $ that matches the \textit{s}-$p_{x}$ band gap,
resulting in a two-photon Raman coupling between these two bands. All lasers
are far-detuned to avoid heating from spontaneous emission. The overall
time-dependent lattice potential can be written as
\begin{equation}
V(\bm{r},t)=\sum\limits_{\eta =x,y,z}V_{0}\cos ^{2}(k_{L}\eta
)+V_{x}^{\prime }\cos ^{2}\Big(k_{L}x+\frac{\omega t}{2}\Big)~,  \label{EqVt}
\end{equation}%
where $V_{0}$ and $V_{x}^{\prime }$ are the static and moving lattice
depths, $k_{L}=\pi /a$ with the lattice constant $a$.

We consider a large static lattice $V_{0}$, but a weak moving lattice $%
V_{x}^{\prime }$ (i.e., $V_{x}^{\prime }\ll V_{0}$); therefore only on-site
and nearest-neighbor tunnelings need be considered and the total wave
function $|\Psi \rangle $ can be expanded in terms of the static lattice
Wannier functions $|\Psi \rangle =\sum\nolimits_{j}c_{sj}|s^{j}\rangle
+c_{p_{x}j}|p_{x}^{j}\rangle $, where $j$ is the site index in the $x$
direction. $|s^{j}\rangle $ and $|p_{x}^{j}\rangle $ are the $s$-band and $%
p_{x}$-band Wannier functions at the $j$-th lattice site, $c_{sj}$ and $%
c_{p_{x}j}$ are their annihilation operators, respectively. Along the other
two directions, $p$-band is not coupled and only \textit{s}-band is
considered and their related indices are neglected here for simplicity.

Under the Wannier basis, we can derive the single-particle tight-binding
Hamiltonian, where the time dependence in the coupling between different
orbits could be further eliminated using the rotating wave approximation
\cite{sm}, similar to the well-known two level Rabi oscillation. The
difference from the Rabi oscillation is that the two levels here (\textit{s}
and $p_{x}$ bands) have different band dispersions. Physically, there are
three types of possible couplings between \textit{s} and $p_{x}$ bands, as
illustrated in Fig.~\ref{fig-illustrate}(a), with the coupling strengths
given by $\Omega =\frac{V_{x}^{\prime }}{4}\langle s^{i}|\sin
(2k_{L}x)|p_{x}^{i}\rangle $, $\alpha =\frac{V_{x}^{\prime }}{2}\langle
s^{i}|\cos (2k_{L}x)|p_{x}^{i+1}\rangle $, and $\beta =\frac{V_{x}^{\prime }%
}{2}\langle s^{i}|\sin (2k_{L}x)|p_{x}^{i+1}\rangle $. The first term $%
\Omega $ denotes the coupling of two orbital states at the same site, while
the last two terms $\alpha $ and $\beta $ are the couplings between nearest-neighbor
sites~\cite{Note}. The values of $\Omega $, $\alpha $, and $\beta $
calculated from the Wannier functions are plotted in Fig.~\ref%
{fig-illustrate}(b) (see also Fig. S1 \cite{sm}). $\beta $ is usually small
and not important for the physics discussed here.

The resulting time-independent single-particle Hamiltonian in the momentum
space can be written as
\begin{equation}
H_{0}(\bm{k})=\left(
\begin{array}{cc}
\epsilon _{s}(\bm{k})+h & \Pi (k_{x}) \\
\Pi (k_{x}) & \epsilon _{p}(\bm{k})-h%
\end{array}%
\right)  \label{Ham}
\end{equation}%
under the basis ($c_{s}(\bm{k})$, $c_{p_{x}}(\bm{k})$)$^{T}$, where $\Pi
(k_{x})=\Omega -\alpha \sin (k_{x}a)+\beta \cos (k_{x}a)$, $\epsilon _{s}(%
\bm{k})=-2t_{s}[\cos (k_{x}a)+\cos (k_{y}a)+\cos (k_{z}a)]-\mu $ and $%
\epsilon _{p}(\bm{k})=2t_{p}\cos (k_{x}a)-2t_{s}[\cos (k_{y}a)+\cos
(k_{z}a)]-\mu $. $t_{s}$ and $t_{p}$ are the nearest-neighbor tunneling
amplitudes for atoms in the $s$- and $p_{x}$-orbital states, respectively. $%
2h$ is the energy difference between $\omega $ and the band gap $\Delta _{g}$%
. $\mu $ is the chemical potential. Note that $\alpha \sin (k_{x}a)\sigma
_{x}$ corresponds to the band-pseudospin-momentum coupling.

In the absence of $\alpha $, $H_{0}(-\bm{k})=H_{0}(\bm{k})$, revealing that
the single-particle Hamiltonian is symmetric under inversion transformation.
This inversion symmetry is broken when $\alpha $ and $\Omega $ coexist. A
typical single-particle band structure, which is asymmetric along $k_{x}$%
-axis, is shown in Fig. \ref{fig-illustrate}(c). Here we just show the Fermi
surface in the $k_{y,z}=0$ plane. The Fermi surface is still symmetric along
$k_{y}$ and $k_{z}$ directions. The orbital and hyperfine-spin degrees of
freedom of the atoms are independent; therefore the coupling between
different orbital states does not break the spin degeneracy and the hybrid
bands are spin balanced at any $\bm{k}$ point.

\textit{Pairing Hamiltonian}: Consider a spin-1/2 Fermi gas with equal spin
populations loaded on such an asymmetric orbital band. The dominant on-site
atom-atom interaction between opposite spins can be made attractive via
Feshbach resonance, similar as regular two component Fermi gases \cite%
{Zwierlein2004}. As a good approximation, the on-site atom-atom interaction
can take the same form as the time-independent static system \cite{sm}. In
the momentum space, the inter- and intra-band interaction term can be
written as ${H}_{\mathrm{int}}=-\sum_{\mu \nu }g_{\mu \nu }c_{\downarrow \mu
}^{\dag }(\bm{k}_{1})c_{\uparrow \nu }^{\dag }(\bm{k}_{2})c_{\uparrow \nu }(%
\bm{k}_{3})c_{\downarrow \mu }(\bm{k}_{4})$, where $\bm{k}_{1}+\bm{k}_{2}=%
\bm{k}_{3}+\bm{k}_{4}$ due to the momentum conservation for the two-body
scattering process. $\mu $ and $\nu $ denote the orbital states of two
spins. $g_{\mu \nu }=g\int dx|w_{\mu }(x)|^{2}|w_{\nu }(x)|^{2}$ is the
interaction coefficient for two atoms in two orbital states (labeled by $%
\mu $ and $\nu $), and $g$ is the two-body interaction strength in free
space. To compare the strengths of the interactions between two orbital
states, we approximate the lattice potential at each site by a harmonic
trap, which is a good approximation when the static lattice is not very
weak. The relative ratio of the interaction strength is found to be $%
g_{ss}:g_{sp}:g_{ps}:g_{pp}=1:0.5:0.5:0.75$ \cite{sm}. Hereafter, we denote $%
g_{ss}=U$.

Under the mean-field approximation, we can rewrite the interaction term with
the effective pairing between atoms. Because the inversion symmetry is
broken for the single-particle Hamiltonian, the system may favor Cooper
pairing with a finite center-of-mass momentum between two fermions of
opposite spins. For simplicity, the chemical potential is chosen
appropriately where there is only one simple Fermi surface (see Fig. \ref%
{fig-illustrate}(c)),{\ therefore we could consider a plane-wave FF-type
inter- and intra-band pairing $\Delta _{\mu \nu }(\mathbf{x})=\Delta _{\mu
\nu }e^{i\mathbf{Q}\cdot \mathbf{x}}$}, similar as that in spin-orbit
coupled system \cite%
{Zheng2013,Wu2013,Dong2013,topoFF1,topoFF2,topoFF3,topoFF4}{\textbf{.} Here $%
\Delta _{\mu \nu }=g_{\mu \nu }\langle c_{\uparrow \mu }(\bm{Q}/2+\bm{k}%
)c_{\downarrow \nu }(\bm{Q}/2-\bm{k})\rangle $ denotes the amplitude of the
s-wave order parameter between two orbital states $\mu $ and $\nu $, and the
FF vector $\bm{Q}=(Q,0,0)$ is the Cooper pairing momentum which is along the
moving lattice direction.} Note that the effective pairing on the asymmetric
Fermi surface (Fig. \ref{fig-illustrate}c) {could be }$\bm{k}${-dependent
(i.e., with non-\textit{s}-wave components) due to the }$\bm{k}$-dependent
hybridization coefficients (determined by the eigenfunction of the
Hamiltonian (\ref{Ham})) of two orbital bands for the asymmetric Fermi
surface \cite{Zhang2008}.{\ }In the basis of spinor ($\Psi (\bm{Q}/2+\bm{k})$%
, $\Psi ^{\ast }(\bm{Q}/2-\bm{k})$)$^{T}$ with $\Psi =(c_{\uparrow
s},c_{\downarrow s},c_{\uparrow p_{x}},c_{\downarrow p_{x}})^{T}$, the
Bogliubov-de Gennes (BdG) Hamiltonian can be written as
\begin{equation}
H_{\mathrm{BdG}}(\bm{k})=\left(
\begin{array}{cc}
H_{0}(\frac{\bm{Q}}{2}+\bm{k})\otimes \sigma _{0} & \Delta _{4\times 4} \\
\Delta _{4\times 4}^{\dag } & -H_{0}(\frac{\bm{Q}}{2}-\bm{k})\otimes \sigma
_{0}%
\end{array}%
\right) ~,  \label{bdg}
\end{equation}%
where $\sigma _{i}$ ($i=x,y,z,0$) are the Pauli matrices,
\begin{equation}
\Delta _{4\times 4}=\left(
\begin{array}{cc}
\Delta _{ss} & -\Delta _{sp} \\
\Delta _{ps} & \Delta _{pp}%
\end{array}%
\right) \otimes (-i\sigma _{y})~.
\end{equation}

For each set of system parameters $(V_{0},V_{x}^{\prime },U)$, the
corresponding parameters $\Omega $, $\alpha $, $\beta $, $t_{s}$, $t_{p}$ in
the BdG Hamiltonian (\ref{bdg}) are calculated from the Wannier functions,
from which the order parameter amplitude $\Delta _{\mu \nu }$ and the FF
vector $\mathbf{Q}$ are simultaneously obtained by minimizing the
thermodynamic potential. When $\Delta _{\mu \nu }\neq 0$ and $Q\neq 0$, the
system is in an FF phase. When $\Delta _{\mu \nu }\neq 0$, $Q=0$, the system
is in a BCS phase. Otherwise, the system is a normal gas.

\begin{figure}[t]
\centering\includegraphics[width=0.48\textwidth]{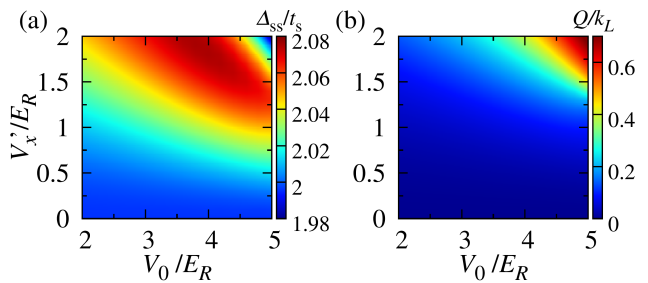}
\caption{(Color online) Phase diagrams of FF superfluids. The color
describes the amplitude of (a) the order parameter $\Delta _{ss}$ and (b)
the FF vector $Q$. Other parameters are $U=6.0t_{s}$, $\protect\mu %
=10.0t_{s} $, $h=8.0t_{s}$. }
\label{fig-omega-alpha}
\end{figure}

\textit{Phase diagrams}: In Fig. \ref{fig-omega-alpha} we plot the
intra-orbital order parameter $\Delta _{ss}$ and the Cooper pairing momentum
$Q$ with respect to the static and moving lattice depths $V_{0}$, $%
V_{x}^{\prime }$. $\Delta _{sp}$ and $\Delta _{pp}$ are much smaller than $%
\Delta _{ss}$ \cite{sm}, which is ascribed to the initial dominant
populations of the $s$-orbital band at the position of the chemical
potential. The $s$- and $p_{x}$-orbital band tunneling and coupling
parameters ($t_{s}$, $t_{p}$, $\Omega $, $\alpha $) depend on the static
lattice depth $V_{0}$ implicitly; therefore $\Delta _{ss}$ does not change
monotonically. However, $\Omega $ and $\alpha $ depend on $V_{x}^{\prime }$
linearly, which directly determine the single-particle band structure;
therefore $Q$ increases with increasing $V_{x}^{\prime }$. Because the
coupling between $s$- and $p_{x}$-orbital states does not depend on spins
(the internal states) of atoms, $\Omega $, $\alpha $ and $\beta $ modify the
energy dispersion in the same way for the two spins, leading to the spin
degenerate asymmetric Fermi surface as shown in Fig. \ref{fig-illustrate}%
(c). Such spin-balanced asymmetric Fermi surface has little effect on
suppressing the order parameter, in contrast to the strong suppression of
the finite momentum pairing order induced by an external Zeeman field.
Therefore, $\Delta _{ss}$ is large and does not change much in the whole
parameter region. $Q$ is proportional to both $\Omega $ and $\alpha $ as
shown in Fig. \ref{fig-omega-alpha}(b). When $V_{x}^{\prime }=0$, all the
coupling coefficients vanish and the band inversion symmetry is preserved;
thus the superfluid becomes a conventional BCS state.

\begin{figure}[b]
\centering\includegraphics[width=0.48\textwidth]{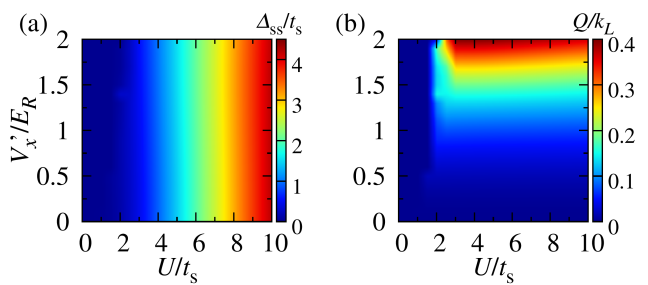}
\caption{(Color online) Phase diagrams in the BCS-BEC crossover. The color
describes the amplitude of (a) the order parameter $\Delta _{ss}$ and (b)
the FF vector $Q$. $V_{0}=4E_{R}$ with $t_{s}=0.0855E_{R}$. Other parameters
are $\protect\mu =10.0t_{s}$, $h=8.0t_{s}$.}
\label{fig-crossover}
\end{figure}

When the on-site interaction $U$ is tuned by changing the \textit{s}-wave
scattering length through Feshbach resonance, the system undergoes a BCS-BEC
crossover. BCS-BEC crossover physics of Fermi gases has been widely studied
in free space and in lattices \cite{crossover1,crossover2}. Here, we present
the phase diagram in the $U$-$V_{x}^{\prime }$ plane in Fig. \ref%
{fig-crossover}. From Fig. \ref{fig-crossover}(a), we see $\Delta _{ss}$ is
mainly determined by the $U$ and changes only slightly with the increase of
moving lattice depth $V_{x}^{\prime }$. FF states with large $Q$ exist in a
large parameter region which dominates when the moving depth is large. In
the weak and medium interaction regimes, the order parameters are small and
the ground state is mainly governed by the single-particle Hamiltonian.
Therefore $Q$ could be significant because of its sensitivity to the single-particle dispersion.
In the very strong interaction regime, the fermions
form tightly bound molecules and the influence of the asymmetric energy
dispersion on Cooper pairs is negligible. Therefore, $Q$ gradually decreases
for a large $U$ and the ground state eventually becomes a BCS state.

\begin{figure}[t]
\centering\includegraphics[width=0.48\textwidth]{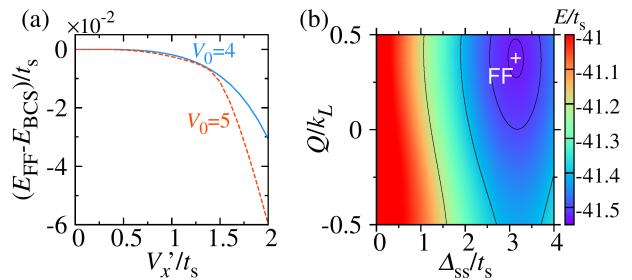}
\caption{(Color online) (a) The thermodynamic potential difference of the FF
states and the possible BCS state as a function of $V_{x}^{\prime }$ for $%
V_{0}=4E_{R}$ (blue solid line, for which $t_{s}=0.0855E_{R}$) and $%
V_{0}=5E_{R}$ (orange dashed line, for which $t_{s}=0.0658E_{R}$). (b) The
contour plot of the thermodynamic potential $E$ in the $\Delta _{ss}$-$Q$
plane for $V_{0}=4.0E_{R}$, $V_{x}^{\prime }=2.0E_{R}$. The cross symbol
corresponds to the self-consistent solution $\Delta _{ss}=3.14t_{s}$, $%
Q=0.377k_{L}$. Other parameters are $U=8.0t_{s}$, $\protect\mu =10.0t_{s}$, $%
h=8.0t_{s}$.}
\label{fig-contourmap}
\end{figure}

\textit{Stability of FF superfluids}: The stability of the FF superfluid may
be characterized by the thermodynamic potential difference $E_{\mathrm{FF}%
}-E_{\mathrm{BCS}}$ between the FF ground state and the possible BCS excited
state (by enforcing $Q=0$), which is shown in Fig. \ref{fig-contourmap}(a).
The larger $|E_{\mathrm{FF}}-E_{\mathrm{BCS}}|$, the FF state is more
stable. When $V_{x}^{\prime }=0$, the inversion symmetry is preserved and
the FF superfluid becomes the BCS state, therefore $E_{\mathrm{FF}}=E_{%
\mathrm{BCS}}$. With the increasing $V_{x}^{\prime }$, $E_{\mathrm{FF}}-E_{%
\mathrm{BCS}}$ becomes negative, indicating that the asymmetric energy
dispersion favors FF superfluids. In Fig. \ref{fig-contourmap}(b), we plot
the thermodynamic potential $E$ in the $\Delta _{ss}$-$Q$ plane for the
premium values of $\Delta _{sp}$ and $\Delta _{pp}$ that minimize the total
energy, which shows that the FF state is indeed the global minimum of the
thermodynamic potential.

We also calculate the energy difference of the FF states and BCS states in a
3D lattice when only an external Zeeman field is applied to break the spin
degeneracy. We find that the energy difference in our spin-balanced model ($%
\sim 3\times 10^{-2}t_{s}$) is much larger than that in the spin-imbalanced
system ($\sim 10^{-3}t_{s}$) with the same interaction strength. Moreover,
the FF states only exist in a very narrow Zeeman field parameter region ($%
\sim 10^{-2}t_{s}$) in the spin-imbalanced schemes and thus it is hard to
find their signature experimentally. In contrast, the FF superfluids in our
spin-balanced system induced by the driven optical lattices exist in almost
the whole parameter region.

$\Omega $ and $\alpha $ are proportional to the moving lattice depth $%
V_{x}^{\prime }$, and can be tuned in a wide parameter range to achieve an
extremely asymmetric energy dispersion. In Fig. \ref{fig-contourmap}(a), we
see $E_{\mathrm{FF}}-E_{\mathrm{BCS}}$ decreases sharply when $V_{x}^{\prime
}$ is large. Therefore with a larger $V_{x}^{\prime }$, $|E_{\mathrm{FF}}-E_{%
\mathrm{BCS}}|$ may be much larger than that shown in Fig. \ref%
{fig-contourmap}(a), which is generally impossible in the spin-imbalanced
Fermi gases. This advantage, together with the large parameter region for FF
states, make our proposed spin-balanced Fermi gas experimentally more
feasible for observing FF superfluids than the spin-imbalanced systems.

\textit{Experimental observation}: The proposed FF superfluids can be
realized with different types of fermionic atoms, such as $^{40}$K and $^{6}$%
Li. In the following, we illustrate the experimental setup and observation
using $^{40}$K. The ultracold $^{40}$K gas with a spin-balanced mixture of
internal states $|F,m_{F}\rangle =|9/2,-9/2\rangle $ and $|9/2,-7/2\rangle $
\cite{Esslinger2014} is trapped in a 3D static optical lattice created by
counter-propagating far-detuned lasers with wavelength $\lambda =1064\text{
nm}$ that defines the wavevector $k_{L}=2\pi /\lambda $ and the recoil
energy $E_{R}=\hbar ^{2}k_{L}^{2}/2m=2\pi \hbar \times 4.5\text{ kHz}$. The
lowest two orbital bands, $s$- and $p_{x}$-orbital, have a gap $\Delta
_{g}\approx 2.6E_{R}$ when the static lattice depth is tuned as $%
V_{0}=3.0E_{R}$. The 1D moving lattice, created by another two
counter-propagating lasers with a slight frequency difference of $\omega
\sim \Delta _{g}/\hbar $, can be tuned to have a lattice depth of $%
V_{x}^{\prime }$ $=0.1\sim 0.8E_{R}$. With these parameters, the resulting
coupling strengths have a range of $\Omega $ $=0.24\sim 1.88t_{s}$ and $%
\alpha $ $=0.19\sim 1.56t_{s}$ ($t_{s}=0.111E_{R}$). The maximum value of FF
momentum $Q$ could be as large as $0.3k_{L}$ and the corresponding order
parameter $\Delta \sim t_{s}$. For $^{6}$Li atoms, the recoil energy becomes
$E_{R}=2\pi \hbar \times 30.0$ kHz for the same $1064$ nm\ lasers and the
relation of all other parameters with $E_{R}$ are the same as $^{40}$K.
Signatures of FF superfluids can be captured by the atom shot noise \cite%
{Greiner}, or the sound speed measurement~\cite{Ketterle,Thomas,Xu2014}.

\textit{Discussion}: We note that recently similar moving lattices have been
used to couple \textit{s}-orbital states of neighboring sites of a tilted
optical lattice for the simulation of the magnetic Hofstadter model with a
BEC~\cite{Bloch2013,Ketterle2013}. In a more general context, similar time
periodic modulation of the lattice to couple different Bloch bands (\textit{%
e.g.}, shaken lattices), known as \textquotedblleft Floquet engineering"~%
\cite%
{Goldman2014a,Goldman2014b,Rigol2014,Lindner,Jiang2011,Wang2013,tunableBEC},
has been investigated extensively in experiments, leading to the observation
of various phenomena such as the paramagnetic to ferromagnetic transition
(bosons)~\cite{Chin2013} and the simulation of topological Haldane model
(fermions)~\cite{Esslinger2014}, where the atomic spin states are
irrelevant. However, the effects of \textit{s}-wave interaction between two
spins of the Fermi gas has not been well explored and our proposed FF
superfluids showcase the rich quantum phases that may be generated by the
\textit{s}-wave two-body interactions in such Floquet systems. Furthermore,
the proposed band-pseudospin-momentum coupling in optical lattices may open
a new avenue for exploring exotic spin-orbit coupling physics.

\begin{acknowledgments}
\textit{Acknowledgements:} We thank Fan Zhang for helpful discussions. C. Qu
and C. Zhang are supported by ARO (W911NF-12-1-0334) and AFOSR
(FA9550-13-1-0045). Z. Zheng and X. Zou are supported by National Natural
Science Foundation of China (Grants No. 11074244 and No. 11274295), and
National 973 Fundamental Research Program (No. 2011cba00200).

\textit{Note added}: Recently, the
modification of the band structure by the moving lattice has been observed
experimentally with a BEC \cite{Amin}.
\end{acknowledgments}


\begin{thebibliography}{99}
\bibitem{Note} Because of different parities of the Wannier functions, the
coupling between two orbital states at the same site mediated by $\cos
(2k_{L}x)$ vanishes. The moving lattice also induces couplings between the
same orbital states ($\mu =\nu $). These terms can also be eliminated by
changing to the rotating frame coordinates after a unitary transformation
\cite{sm}.
\end{thebibliography}
\end{document}